\documentclass[final,2p,times]{elsarticle}
\usepackage[margin=0.9in]{geometry}
\usepackage{graphicx}
\usepackage{amssymb}
\usepackage{amsmath}
\usepackage{lineno}
\usepackage{placeins}
\usepackage{xcolor}
\usepackage{xspace}
\usepackage{hyperref}
\usepackage[scientific-notation=true]{siunitx}
\journal{}

\usepackage{booktabs}

\newcommand{\todo}[1]{}
\renewcommand{\todo}[1]{{\color{red} TODO: {#1}}}

\newcommand*{\Rinv}{\ensuremath{R_{\textrm{inv}}}}

\newcommand*{\RinvR}{\ensuremath{R^{\textrm{Ratio}}_{\textrm{inv}}}}
\newcommand*{\RinvBF}{\ensuremath{R^{\textrm{BF}}_{\textrm{inv}}}}

\newcommand*{\minphi}{\ensuremath{\phi_{\textrm{MET},\textrm{closest-jet}}}}

\def \HT {\ensuremath{H_{\text{T}}}\xspace}

\def \bjets {\ensuremath{b\text{-jets}}\xspace}
\def \pT {\ensuremath{p_{\mathrm{T}}}\xspace}
\def \met {\ensuremath{p_{\text{T}}^{\text{miss}}}\xspace}

\makeatletter
\def\ps@pprintTitle{%
 \let\@oddhead\@empty
 \let\@evenhead\@empty
 \def\@oddfoot{\centerline{\thepage}}%
 \let\@evenfoot\@oddfoot}
\makeatother

\begin{document}

\begin{frontmatter}

\title{Probing the sensitivity of semi-visible jets to current LHC measurements\\ using the CONTUR toolkit}

\author[labelAB]{Andy~Buckley}
\author[labelJB]{Jon~Butterworth}
\author[labelLC]{Louie~Corpe }
\author[labelUofM]{Caterina~Doglioni}
\author[labelDK]{\\Deepak~Kar \footnote{deepak.kar@cern.ch}}
\author[labelCP]{Clarisse~Prat \footnote{clarisse.prat@cern.ch}}
\author[labelUofM]{Sukanya~Sinha}
\author[labelUofM]{Danielle~Wilson-Edwards}

\address[labelAB]{Department of Physics \& Astronomy, University of Glasgow, Glasgow, United Kingdom} 

\address[labelJB]{Department of Physics \& Astronomy, University College London, London, United Kingdom} 

\address[labelUofM]{School of Physics and Astronomy, University of Manchester, Manchester, United Kingdom} 

\address[labelLC]{Université Clermont Auvergne, CNRS/IN2P3, LPCA, 63000 Clermont-Ferrand, France} 

\address[labelSK]{Institute of Physics, NAWI Graz, University of Graz, Graz, Austria} 

\address[labelDK]{School of Physics, 
University of Witwatersrand, Johannesburg, South Africa, and \\ 
Royal Society Wolfson Visiting Fellow at the University of Glasgow, United Kingdom}

\address[labelCP]{School of Physics, 
University of Witwatersrand, Johannesburg, South Africa}

\begin{abstract}
Semi-visible jets arise from a hypothetical, strongly interacting ``dark sector'' -- a dark counterpart of quantum chromodynamics whose partial decays back to Standard Model particles introduce new types of collider BSM signature. CMS and ATLAS have have searched for semi-visible jets in the resonant and non-resonant production modes and set constraints on mediator mass values. In this work, indirect constraints on various model parameters, such as dark hadron masses and coupling strengths, are explored using LHC measurements.
\end{abstract}


\end{frontmatter}


\section{Introduction}
\label{sec:intro}

A long-standing mystery of fundamental physics is the nature of dark matter (DM), a type of matter whose existence is supported by astrophysical and cosmological observations~\cite{Bertone:2004pz} and that has only feeble, if any, interactions with ordinary matter. 
Recent theory proposals postulate that DM may be part of a complex group of particles in a strongly-coupled dark sector~\cite{Schwaller:2015gea,Beauchesne:2018myj,Bernreuther:2019pfb}.
These models give rise to distinctive and not fully explored collider topologies, such as \textit{semi-visible jets} (SVJ)~\cite{Cohen:2015toa,Cohen:2017pzm,Beauchesne:2017yhh}. 
Such SVJs have dark hadrons geometrically inside them.
This results in a topology in which the direction of the missing transverse momentum, termed \met, 
is aligned with the semi-visible jet. 

Experimental results of searches for SVJ have been presented in the resonant-production mode by the CMS collaboration~\cite{CMS-EXO-19-020}, and in the non-resonant mode by the ATLAS collaboration~\cite{ATLAS:2023swa}. 
Recent works have proposed ways to ascertain systematic uncertainties on SVJ production~\cite{Cohen:2020afv}, test the use of jet substructure or similar observables~\cite{Park:2017rfb,10.21468/SciPostPhys.10.4.084,Cohen:2023mya}, or machine-learning methods to understand and discriminate SVJs from Standard Model (SM)jets~\cite{Bernreuther:2020vhm, Canelli:2021aps, Faucett:2022zie, Finke:2022lsu, Pedro:2023sdp, Bhardwaj:2024djv}.
New signatures involving SVJs have also been proposed, such as with light leptons~\cite{Cazzaniga:2022hxl}, tau leptons~\cite{Beauchesne:2022phk}, photons~\cite{Cazzaniga:2024mmv} or new production modes such as with ISR~\cite{Liu:2024rbe}, \bjets~\cite{10.21468/SciPostPhysCore.7.4.071} or from dark glueballs~\cite{Batz:2023zef}. 
The Snowmass Whitepaper~\cite{Albouy:2022cin} and the MITP Colours in Darkness workshop summary report~\cite{Butterworth:2023cgz} have discussed the current status of SVJ modelling in Monte Carlo event-generation frameworks. 

In this work, we describe the parameters and process to generate SVJ samples starting from the current state of the art. We then use the generated samples to re-interpret SM measurements using the Constraints On New Theories Using Rivet (Contur) toolkit~\cite{Butterworth:2016sqg,Buckley:2021neu,Bierlich:2019rhm}, and derive constraints on selected SVJ models and parameter choices.

\section{Semi-visible jets generation}
\label{sec:gen}

A pair of dark quarks are generated via a heavy $Z^{\prime}$ mediator in the $s$-channel, or via a scalar bifundamental mediator in the $t$-channel. 
The $s$-channel model is generated using Pythia\,8~\cite{Sjostrand:2014zea}. 
The $t$-channel model is generated using the UFO in Ref.~\cite{dmsimp} in Madgraph~\cite{Alwall:2014hca, Frederix:2012ps}~\footnote{Here the particle IDs of the DM particles were changed to match the Pythia\,8 Hidden Valley scheme.}, following Ref.~\cite{Cohen:2017pzm}.
In both cases, the dark quarks undergo dark hadronisation using the Pythia\,8 Hidden Valley~\cite{Carloni:2011kk} module, which gives rise to  flavour-diagonal and off-diagonal $\pi_{\mathrm{d}}$ and $\rho_{\mathrm{d}}$ mesons, with spin~0 and~1 respectively.

The Hidden Valley parameters of importance in these models are:
\begin{itemize}
\item the dark QCD confinement scale value ($\Lambda_{D}$), 
\item the number of dark colours ($N_C$), 
\item the number of dark flavours ($N_F$), 
\item the masses of the dark quarks ($M_{q_{D}}$) 
\item the masses of the dark hadrons ($M_{\rho_{D}}$, $M_{\pi_{D}}$).
\end{itemize}

For the semi-visible jets scenario, unstable dark hadrons decay to stable dark hadrons and 
to SM quarks. 
The masses of the dark hadrons were chosen so that the above decays are kinematically possible --  it was however observed that these and other detailed choices made for the dark sector parameters did not influence the final state topology.

The fraction of decays of dark quarks into SM particles is characterised by the $(1-\Rinv)$ parameter. 
\Rinv\ is not an fundamental theory parameter, but we use it as a handle to generate and scan different kinematics for the collider final states. 
In this study, flavour-diagonal $\pi_{\mathrm{d}}$ and $\rho_{\mathrm{d}}$ mesons were promptly decayed into off-diagonal stable dark $\pi_{\mathrm{d}}$ and $\rho_{\mathrm{d}}$ pairs respectively, and
to SM quarks. The branching fraction of the unstable dark mesons to stable dark mesons was set to the value of the \Rinv\ parameter. However for the two-dark-flavours model ($N_F=2$), used in this study, this definition does not lead to the same fraction of dark hadrons in the output, as discussed in more detail in 
\ref{sec:rinv}.

\section{Constraints on semi-visible jets}
\label{sec:contur}

The Constraints On New Theories Using Rivet (Contur) toolkit~\cite{Butterworth:2016sqg,Buckley:2021neu} was used to evaluate the constraints of existing LHC measurements on SVJ signatures. 

Contur uses a large number of the measurements and searches implemented in Rivet~\cite{Bierlich:2019rhm,Bierlich:2024vqo} to constrain a given BSM model.
This approach has been shown to be highly complementary to the direct-search programme~\cite{Buckley:2020wzk,Butterworth:2020vnb, Butterworth:2021jto,Allanach:2021gmj,Altakach:2021lkq}.
The goal of this paper is gain insight from measurements that is complementary to the current experimental results from searches for both $s$- and $t$-channel SVJs. 
Since the SVJ final states are dominated by jets and \met\, measurements in purely hadronic final states are considered for this study. 

In the following, we start from the Simple-NF2 setup mentioned above.
The mass of the dark quark was set to 10~GeV, the flavour-diagonal $\rho_{\mathrm{d}}$ meson mass was fixed as 1.6 times the flavour-diagonal $\pi_{\mathrm{d}}$ mass following Ref.~\cite{Albouy:2022cin}.
The corresponding off-diagonal $\pi_{\mathrm{d}}$ and$\rho_{\mathrm{d}}$ meson masses were set to 0.01~GeV less than the half of the corresponding flavour-diagonal ones.
The number of dark colours was fixed at~3, and the Pythia\,8 HV $\alpha_{\textrm{dark}}$ coupling was chosen to be running at one-loop. 
The dark-QCD confinement scale, $\Lambda_{D}$, value was calculated for each dark meson mass following the formula given in Ref.~\cite{Albouy:2022cin}. 
The lowest allowed \pT of the HV Final State Radiation (FSR) emission (pTMinFSR) was set at 1.1 times $\Lambda_{D}$. 

These choices allowed a relatively simple scan to be performed on the mediator mass, flavour-diagonal $\pi_{\mathrm{d}}$ mass, and input \Rinv\, parameter. 
Additionally, in the $s$-channel generation, a further smaller scan was performed over $N_F$. 
For the $t$-channel study, another free parameter scanned was the strength of the coupling connecting the SM and DM sectors ($\lambda$), which effectively translates into a cross-section scaling~\footnote{The cross-section scales as $\lambda^4$ without having any impact on the kinematic distributions, or on the validity of the model, if the mediator mass is 2.5~TeV or higher.}.
Table~\ref{tab:paramscan} gives a list of parameters scanned over, and the scan ranges. 

\begin{table}[t]
\centering
\begin{tabular}{lrr}
\toprule
Parameter & Range & Comments \\
\midrule
$M_{Z^{\prime}}$ or $M_{med}$ &  1,3,5 TeV    & Both channels  \\
$M_{\pi_{\mathrm{d}}}$ &  10, 50, 100 GeV  & Both channels  \\
\Rinv  & 0.0 -- 1.0  & Both channels \\
$N_F$ &  1, 2, 3   & Only $s$-channel  \\
$\lambda$  &  0.1 - 1   & Only $t$-channel  \\
\hline
$M_{\rho_{\mathrm{d}}}$ & 16, 80, 160~GeV & Function of $M_{\pi_{\mathrm{d}}}$ \\
$\Lambda_{D}$ & 5, 25, 50   & Function of $M_{\pi_{\mathrm{d}}}$ and $M_{\rho_{\mathrm{d}}}$ \\
pTMinFSR  &   5.5, 27.5, 55  & Function of $\Lambda_{D}$  \\
\bottomrule
\end{tabular}
\caption{Parameters and ranges on which the Contur scan was performed. The lower block represents
parameters whose value was set by other parameters from the upper block.}
\label{tab:paramscan}
\end{table}

\FloatBarrier

\subsection{Resonant production mode}

The $s$-channel SVJ signature was generated in Pythia\,8(312), with a $Z^{\prime}$ mediator decaying to a dark-quark pair in the HV module.
The cross-section of this process is very low in the default Pythia\,8 setup, so it has been scaled up to the cross-section used in CMS search~\cite{CMS-EXO-19-020}. 
The cross-section values can be found in Table~\ref{tab:xs-schan}.
As the CMS search has set very strong constraints on this topology, and the cross-section can be highly model-parameter-dependent, we consider this study a proof-of-principle exercise where the sensitivity of measurements on the shapes of the kinematic distributions is explored.

\begin{table}[th]
    \centering
    \begin{tabular}{lrr}
    \toprule
    $M_{Z^{\prime}}$ [TeV] & Generated cross-section [pb] & Scaled cross-section [pb] \\
    \midrule
        1.0 & \num{1.29 e-02}  & 6.69 \\
        1.5 & \num{1.42 e-03}  & 1.88 \\
        2.0 & \num{2.48 e-04} & 0.47 \\
        2.5 & \num{5.68 e-05}  & 0.13 \\
        3.0 & \num{1.49 e-05} & 0.004 \\
    \bottomrule
    \end{tabular}
    \caption{The generated and scaled cross-sections for different $M_{Z^{\prime}}$ values for the $s$-channel process.}
    \label{tab:xs-schan}
\end{table}

The most sensitive analysis for this signature is the ATLAS cross-section measurement of events with missing transverse momentum and jets~\cite{ATLAS:2024vqf}, where only the cross-section distributions were used, rather than their ratios.

\begin{figure}[ht]
  \centering
  \includegraphics[width=0.6\textwidth]{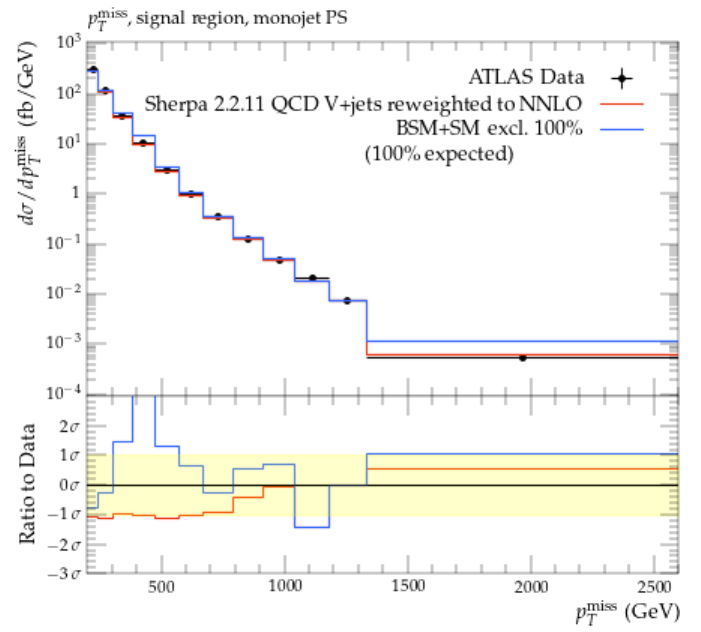}
  \caption{The dominant exclusion histogram for the $s$-channel process
  with $M_{Z^{\prime}} = 1000$~GeV, $M_{\pi_{\mathrm{d}}}=100$~GeV, \Rinv $= 0.5$, and $N_F=2$. 
  The exclusion is driven by the \met distribution measured in ATLAS~\cite{ATLAS:2024vqf} (specifically, Figure~7a).
   The black markers represent the observed data, and the red line shows the SM prediction. 
   The blue line shows the predicted events from the new physics benchmark added to the SM prediction, which in this case results in a significant disagreement with the number of observed events in data.}
  \label{fig:schan-kin}
\end{figure}

In Fig.~\ref{fig:schan-kin} shows the \met distribution for $M_{Z^{\prime}}$ of 1000~GeV, $M_{\pi_{\mathrm{d}}}$ of 100 GeV. The $M_{\pi_{\mathrm{d}}}$ is a key contributor which drives the exclusion values, as will be seen subsequently.
This is for a representative value of \Rinv\ of 0.5 and $N_F=2$.

\begin{figure}[ht]
  \centering
  \includegraphics[width=0.3\textwidth]{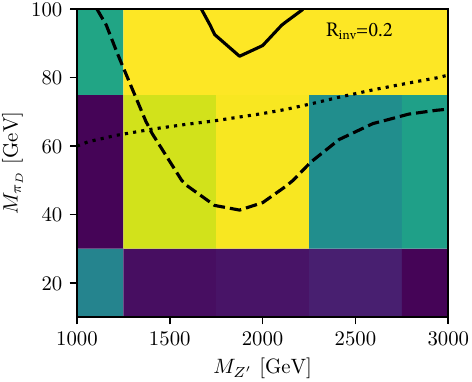}
  \includegraphics[width=0.3\textwidth]{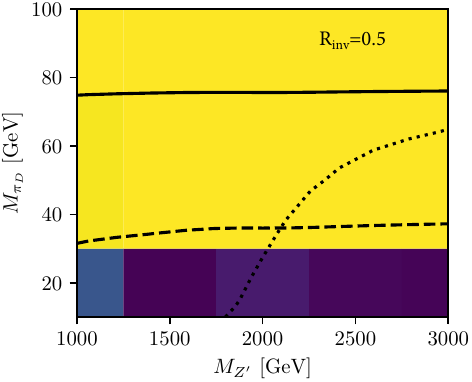}
  \includegraphics[width=0.3\textwidth]{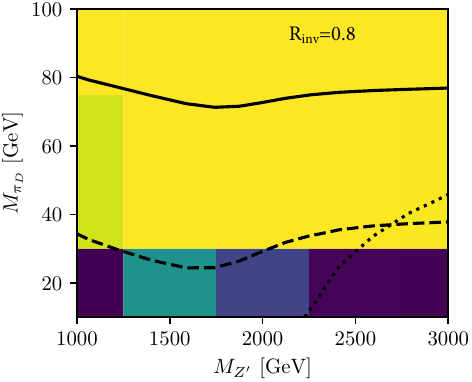}
  \caption{Exclusion contours for the $s$-channel process as a function 
  of $M_{\pi_{\mathrm{d}}}$ and $M_{Z^{\prime}}$ for \Rinv\ values 
  of 0.2, 0.5 and 0.8 are shown respectively.   
  The solid (dotted) lines represent the observed (expected) exclusions at 95\% CL. 
  The dashed line represents the observed 68\% CL exclusion. 
  The area above the curves is excluded.}
  \label{fig:schan-excl1}
\end{figure}

In Fig.~\ref{fig:schan-excl1}, the exclusion of $M_{\pi_{\mathrm{d}}}$ for the specific choice of production cross-sections as a function of $M_{Z^{\prime}}$ for \Rinv\ values of 0.2, 0.5 and 0.8 are shown. 
The binned heatmaps used for the exclusion plot range from dark blue (0~CLs) to yellow (1.0~CLs), and the exclusion contours are drawn from this, where CLs refer to the common limit setting technique of Confidence limits~\cite{Read:2002hq}.
The black solid line is the 95\%-confidence exclusion, the black dashed line is 68\% and the black dotted line is the 95\% expected exclusion.

The main conclusion for this study is that for the specific coupling value of the $Z^{\prime}$ to dark quarks (where 100\% branching fraction for this decay is assumed), measurements place stronger constraints on higher values of the mass of the dark pion $\pi_{\mathrm{d}}$, except at low \Rinv\ and low $M_{Z^{\prime}}$ mass. 

\begin{figure}[ht]
  \centering
  \includegraphics[width=0.3\textwidth]{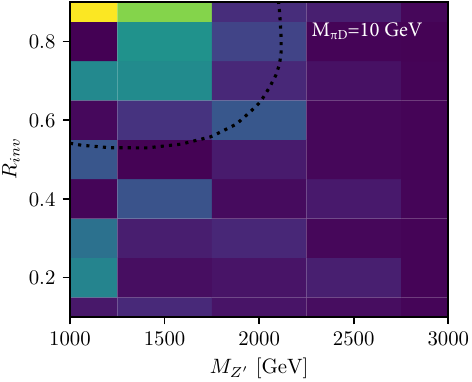}
  \includegraphics[width=0.3\textwidth]{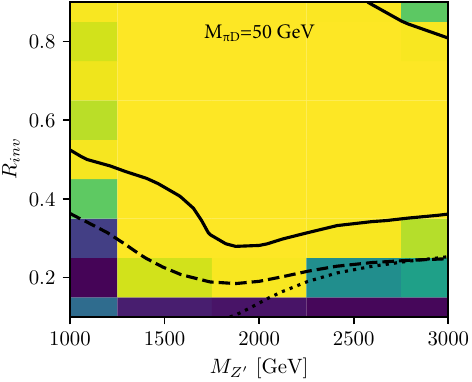}
  \includegraphics[width=0.3\textwidth]{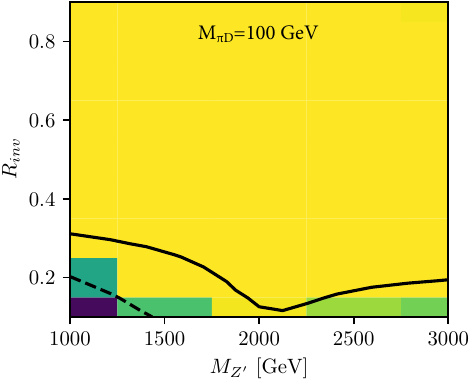}
  \caption{Exclusion contours for the $s$-channel process as a function 
  of \Rinv\ and $M_{Z^{\prime}}$ for $M_{\pi_{\mathrm{d}}}$ values 
  of 10, 50 and 100 GeV are shown.    
  The solid (dotted) lines represent the observed (expected) exclusions at 95\% CL. 
  The dashed line represents the observed 68\% CL exclusion. 
  The area above the curves is excluded.}
  \label{fig:schan-excl2}
\end{figure}

The exclusion plot of \Rinv\ against $M_{Z^{\prime}}$ for different $M_{\pi_{\mathrm{d}}}$ values, shown in Fig.~\ref{fig:schan-excl2} confirms the above conclusion.

\begin{figure}[ht]
  \centering
  \includegraphics[width=0.3\textwidth]{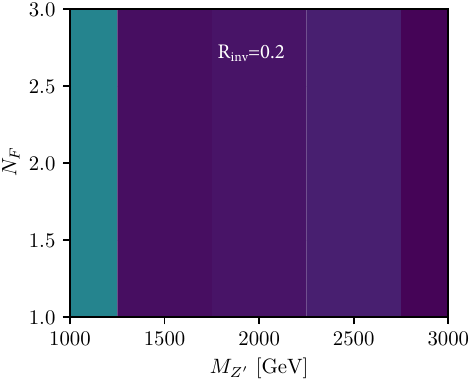}
  \includegraphics[width=0.3\textwidth]{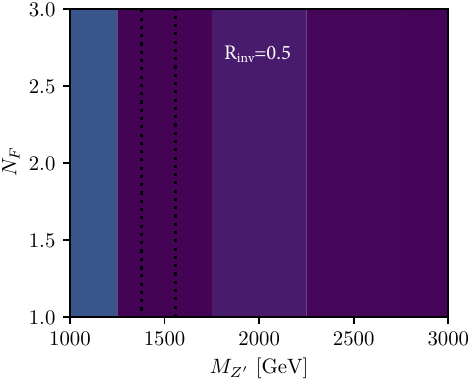} 
  \includegraphics[width=0.3\textwidth]{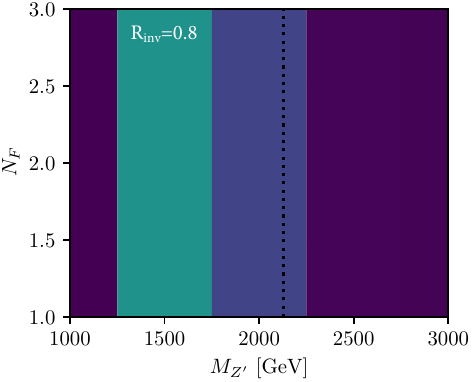} \\
  \includegraphics[width=0.3\textwidth]{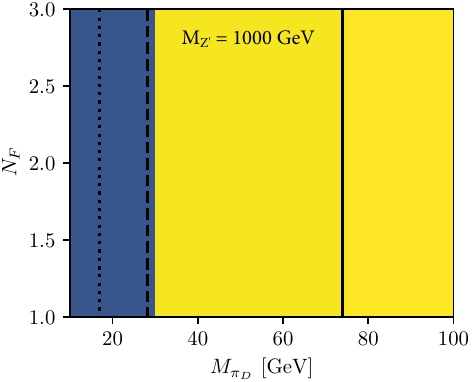}
  \includegraphics[width=0.3\textwidth]{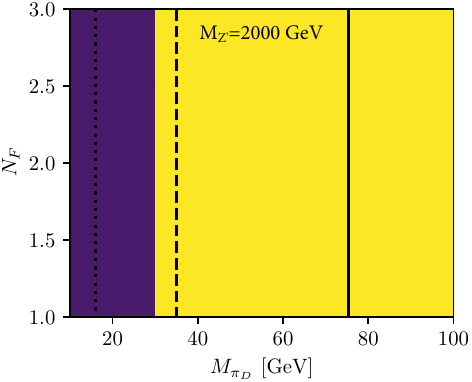}
  \includegraphics[width=0.3\textwidth]{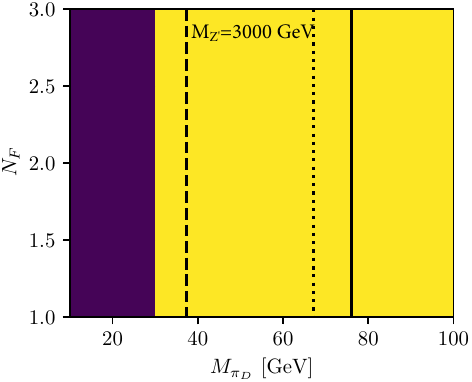} \\
  \caption{Exclusion contours for the $s$-channel process as a function 
  of $N_F$ and \Rinv\ or $M_{Z^{\prime}}$ are shown.    
  The solid (dotted) lines represent the observed (expected) exclusions at 95\% CL. 
  The dashed line represents the observed 68\% CL exclusion. 
  The area above the curves is excluded.}
  \label{fig:schan-excl3}
\end{figure}

The above figures were all shown for $N_F=2$. 
An additional test was performed to check if higher or lower values of $N_F$ would be constrained. 
This can be seen in Fig.~\ref{fig:schan-excl3}, initially for a fixed $M_{\pi_{\mathrm{d}}}$ of 10~GeV varying \Rinv, and then for a fixed \Rinv\ by varying $M_{Z^{\prime}}$. 
In both cases it can be noticed that the sensitivity is dependent on the $\pi_{\mathrm{d}}$ and $Z^{\prime}$ masses, as seen above, while the exclusions are not sensitive to small changes in $N_F$. 
The CMS result, using a $Z^{\prime}$ with a universal coupling of 0.25 to the SM quarks, excluded $M_{Z^{\prime}}$ masses of 1.5-4.0~TeV at 95\% confidence level, depending on the other signal-model parameters.

\FloatBarrier

\subsection{Non-resonant production mode}

The $t$-channel SVJ signature was generated as suggested in Ref.~\cite{dmsimp}, using Madgraph5 to calculate matrix elements (ME) with up to two extra partons at the leading order.
The MLM~\cite{Mangano:2006rw} jet matching scheme, with the matching parameter set to 100~GeV, was employed while showering was done in Pythia\,8(306)~\footnote{
Recent updates to the Pythia\,8 HV module have introduced more rigorous tracking of HV quantum numbers. 
However, these updates are incompatible with the current $t$-channel UFOs, as they do not include particles carrying HV quantum numbers.}. 
Table~\ref{tab:xs-tchan} shows the cross-section for the different mediator masses, for the specific coupling strength choice of $lambda$ of unity. 
An alternative to this process uses the exclusive Pythia\,8 $t$-channel production via $gg$ fusion decaying to a HV partner quark pair, but the cross-section of \num{5.4e-08} pb was found to be too low for current measurements to have any sensitivity to this production mode. 

\begin{table}[h]
    \centering
    \begin{tabular}{lr}
    \toprule
    $M_{med}$ [TeV] & Generated cross-section [pb]\\
    \midrule
        1.0 & 2.11\\
        1.5 & 0.44\\
        2.0 & 0.14\\
        2.5 & 0.006 \\
        3.0 & 0.003 \\
    \bottomrule
    \end{tabular}
    \caption{The generated and scaled cross-sections for different $M_{med}$ values for the $t$-channel process for $\lambda = 1$.}
    \label{tab:xs-tchan}
\end{table}

\begin{figure}[ht]
  \centering
  \includegraphics[width=0.6\textwidth]{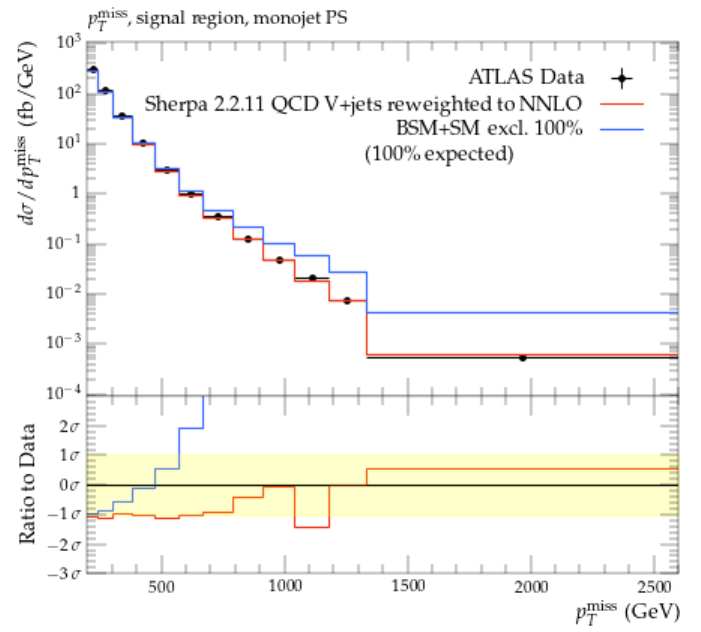}
  \caption{The dominant exclusion histogram for the $t$-channel process
  with $M_{med} = 1000$~GeV, $M_{\pi_{\mathrm{d}}}=50$ GeV, \Rinv $= 5$ and $\lambda=1$. 
  The exclusion comes from the \met distribution measured in ATLAS~\cite{ATLAS:2024vqf},
  specifically, Figure~7a of that paper. The black markers represent the observed data, and the     
  red line shows the SM prediction. The blue line shows the SM prediction summed with the new 
  physics prediction, which in this case results in a significant disagreement with the 
  observation.}
  \label{fig:tchan-kin}
\end{figure}

The most sensitive analysis for the $t$-channel signature is again the ATLAS cross-section measurement of missing transverse momentum and jet events~\cite{ATLAS:2024vqf}. 
Fig.~\ref{fig:tchan-kin} confirms the above observations in showing the \met distribution for $M_{\pi_{\mathrm{d}}}$ of 50~GeV and $\lambda$ of 2.5, where the latter drives the exclusion, as will be seen subsequently. 
This is shown for a representative value of $\Rinv = 0.5$.

\begin{figure}[ht]
  \centering
  \includegraphics[width=0.3\textwidth]{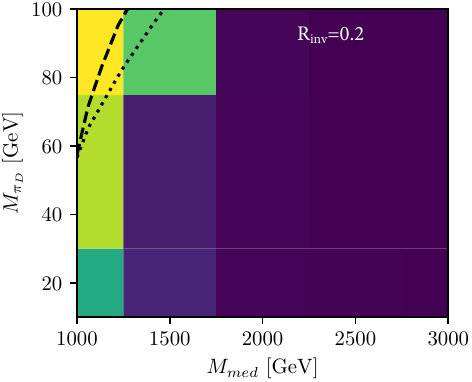}
  \includegraphics[width=0.3\textwidth]{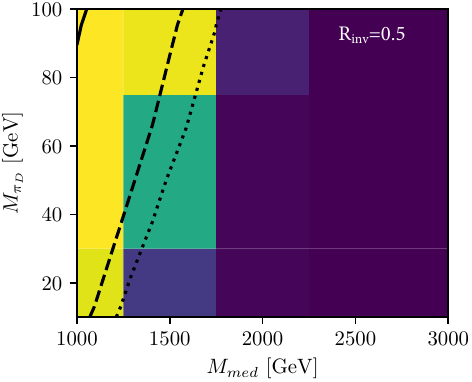}
  \includegraphics[width=0.3\textwidth]{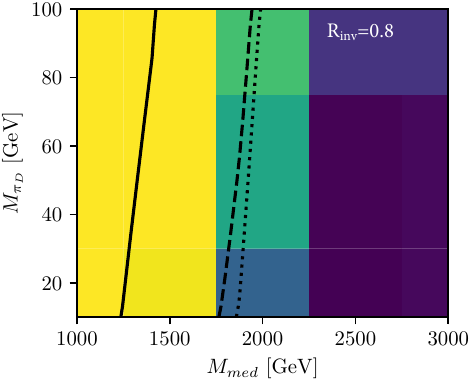}
  \caption{Exclusion contours for the $t$-channel process as a function 
  of $M_{\pi_{\mathrm{d}}}$ and $M_{med}$ for \Rinv\ values 
  of 0.2, 0.5 and 0.8 are shown.    
  The solid (dotted) lines represent the observed (expected) exclusions at 95\% CL. 
  The dashed line represents the observed 68\% CL exclusion. 
  The area above the curves is excluded.}
  \label{fig:tchan-excl1}
\end{figure}

\begin{figure}[ht]
  \centering
  \includegraphics[width=0.3\textwidth]{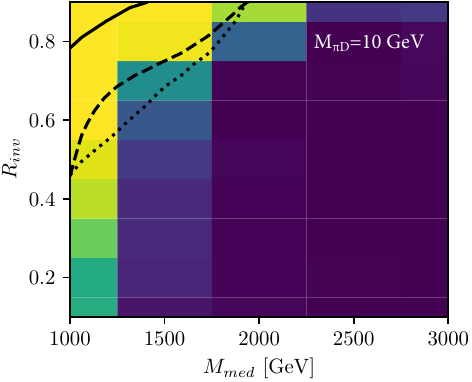}
  \includegraphics[width=0.3\textwidth]{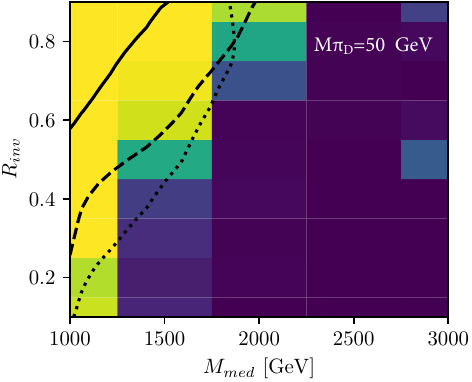}
  \includegraphics[width=0.3\textwidth]{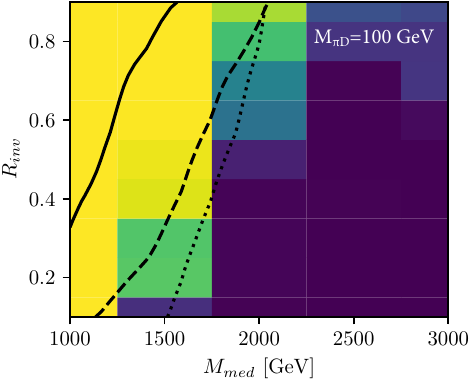}
  \caption{Exclusion contours for the $t$-channel process as a function 
  of \Rinv\ and $M_{med}$ for $M_{\pi_{\mathrm{d}}}$ values 
  of 10, 50 and 100~GeV are shown.       
  The solid (dotted) lines represent the observed (expected) exclusions at 95\% CL. 
  The dashed line represents the observed 68\% CL exclusion. 
  The area above the curves is excluded.}
  \label{fig:tchan-excl2}
\end{figure}

In Fig.~\ref{fig:tchan-excl1}, the exclusion for $M_{\pi_{\mathrm{d}}}$ as a function of $M_{med}$ for \Rinv\ values of 0.2, 0.5 and 0.8 are shown. 
Unlike the $s$-channel results before, there is some dependence on \Rinv\ for excluded mediator masses, where mediators with masses of almost up to 1.75~TeV are excluded for high \Rinv\ values. 
The exclusion plot of \Rinv\ against $M_{med}$ for different 
$M_{\pi_{\mathrm{d}}}$ values, shown in Fig.~\ref{fig:tchan-excl2} 
confirms that the higher values of $\pi_{\mathrm{d}}$ 
mass are mostly excluded, except at low \Rinv\ and low $M_{med}$ masses. 

\begin{figure}[ht]
  \centering
  \includegraphics[width=0.3\textwidth]{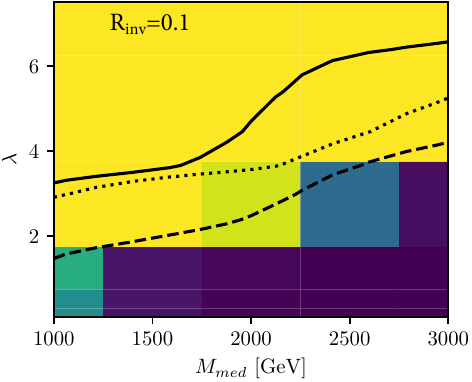}
  \includegraphics[width=0.3\textwidth]{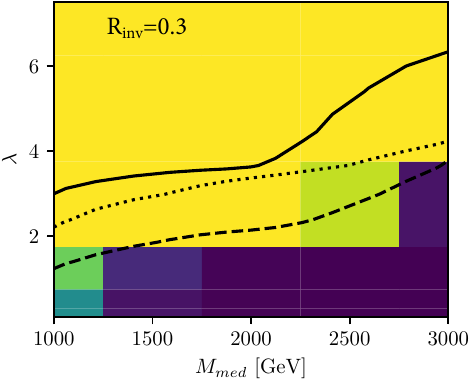} 
  \includegraphics[width=0.3\textwidth]{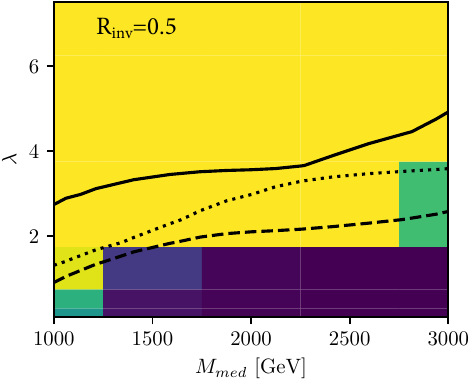} \\
  \includegraphics[width=0.3\textwidth]{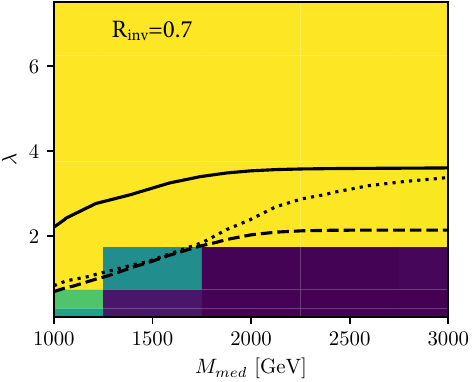}
  \includegraphics[width=0.3\textwidth]{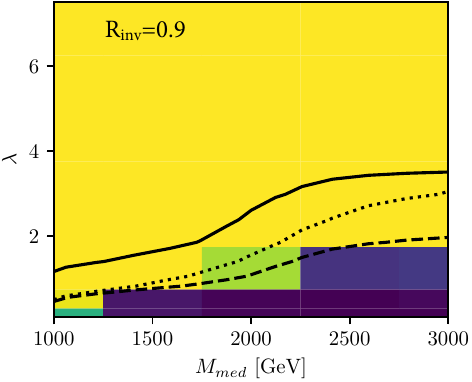}
  \caption{Exclusion contours for the $t$-channel process as a function 
  of $N_F$ and \Rinv\ or $M_{med}$ are shown.    
  The solid (dotted) lines represent the observed (expected) exclusions at 95\% CL. 
  The dashed line represents the observed 68\% CL exclusion. 
  The area above the curves is excluded.}
  \label{fig:tchan-excl3}
\end{figure}

The above results use the cross-section scaling parameter $\lambda$ equal to unity -- for $\lambda$ of 7.5 or higher, the entire parameter grid chosen was excluded. 
In Fig.~\ref{fig:tchan-excl3}, exclusions are shown for a fixed $M_{\pi_{\mathrm{d}}}$ of 10~GeV and varying \Rinv, for $\lambda$ as a function of $M_{med}$.
These results set a stringent limit on the specific production cross-section of these signals. 
The ATLAS search for non-resonant production modes set a limit of up to 2.7~TeV on $M_{med}$ for $\lambda$ of unityand $N_F$ of unity, however a direct comparison with the public experimental result is not possible, given the shift to a physically motivated value of $N_F$ for the present study.

\FloatBarrier

\section{Summary}
\label{sec:summ}

A study covering resonant and non-resonant production mode of semi-visible jets has been presented. The parameter space of benchmark SVJ models generated using the Pythia\,8 HV module are constrained by ATLAS cross-section measurement of events with missing transverse momentum and hadronic jets. 
While these constraints are generally weaker than those set by the CMS and ATLAS searches, they are valid for \met values lower than the ATLAS search in the non-resonant case, showing complementarity with the existing results.


\section*{Acknowledgments}

CP is supported by a SA-CERN excellence bursary.
DK thanks the generous support by Wolfson Foundation and Royal Society to allow him to spend his sabbatical year at the University of Glasgow.
CD, SS and DWE are supported by European Research Council grant REALDARK (Grant Agreement no.~101002463). DWE is also supported by the Science and Technology Facilities Council (STFC).
We would like to thank Suchita Kulkarni, Tim Cohen and Matt Strassler for
useful inputs. 

\bibliographystyle{elsarticle-num.bst}
\bibliography{ref.bib}

\newpage
\appendix
\section{Definition and implementation of invisible fraction}
\label{sec:rinv}

Two different definitions of the \Rinv\ parameter have been used interchangeably in the literature~\footnote{Ref.~\cite{Liu:2024rbe} has also proposed calculating the output \Rinv\ based on the \met fraction in each SVJ, while Ref.~\cite{Kulkarni:2024okx} lets the mass hierarchy of the dark hadrons yield a continuous distribution of an effective output \Rinv\, calculated as ratio of stable dark hadrons over all dark 
hadrons.}:
\begin{enumerate}
\item The fraction of stable dark hadrons relative to all dark hadrons in the event (\RinvR);
\item The event-level fraction of unstable dark hadrons that decay into stable dark hadrons after the dark shower (\RinvBF).
\end{enumerate}

Defining $N_u$ and $N_s$  as the number of unstable and stable dark hadrons, respectively, would imply:
\begin{equation}\label{eqn:rinv}
N_u \RinvBF = \frac{N_s}{2} 
\end{equation}
since each unstable dark hadron decays into a pair of stable dark hadrons.

Now relating \RinvBF\ with \RinvR\, using Eq.~\ref{eqn:rinv}:
\begin{equation} 
	\begin{split}
	 \RinvR & = \frac{N_s}{(N_s+N_u)} \\
	 & = \frac{2 N_u \RinvBF}{(2 N_u  \RinvBF  + N_u)} \\
	 & =  \frac{\RinvBF}{(0.5+\RinvBF}
	\end{split}
\end{equation}

Regardless of event-by-event 
fluctuations, this demonstrates that the two definitions yield different results.
Essentially this means for a lower \RinvBF\, we get a slightly higher \RinvR, and for higher values of \RinvBF, the output ratio is bounded above by 0.67.

For the simplest case of a single dark flavor ($N_F=1$), which is theoretically not well-motivated,
stable dark $\pi$ mesons are exclusively produced through decays. 
In this approach, the unstable dark hadrons correspond to the diagonal ones, while the stable dark hadrons are the off-diagonal ones. 
This approach was adopted in the ATLAS non-resonant search~\cite{ATLAS:2023swa} to maintain a more straightforward and intuitive setup. However, for higher values of $N_F$, stable dark $\pi$ and dark $\rho$ mesons can also be directly produced during dark hadronisation, rather than exclusively through decays. 
The CMS resonant search~\cite{CMS-EXO-19-020}, which used $N_F=2$, treated both diagonal and off-diagonal dark hadrons, whether produced via decay or hadronisation, as unstable. They both decay to generic stable dark matter particles by the same \RinvBF.

\begin{figure}[ht]
\centering
\includegraphics[width=0.45\textwidth]{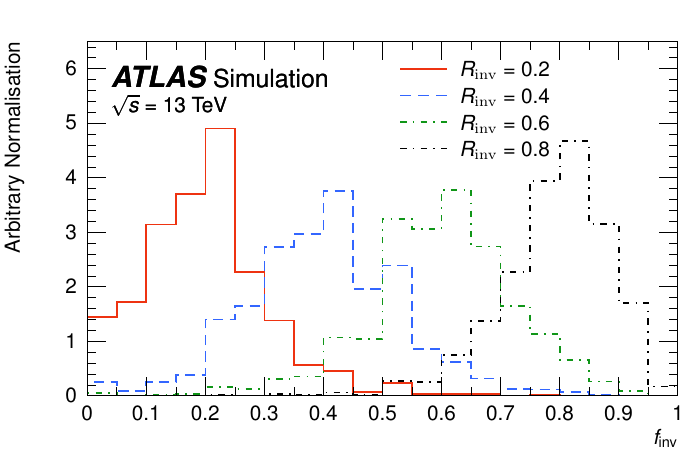}
\caption{The event-level fraction of unstable dark hadrons decaying into stable dark hadrons after the dark shower, from Ref.~\cite{ATLAS:2023swa}. Here the $f_{\textrm{inv}}$ represents the output \RinvBF\ as defined above.}
\label{fig:rinvatlas}
\end{figure}

As the Pythia\,8 run cards consistently use the \RinvBF as \Rinv\, it can be argued that it was intended as an \textit{input parameter}.
The concept of an output \Rinv\ based on dark hadron multiplicities presents challenges, as no experimentally measurable particle level observable can be constructed.
If the output \Rinv\ is also calculated using the \RinvBF\ approach, it trivially matches the input value, as illustrated in Fig.~\ref{fig:rinvatlas}, taken from the ATLAS public result. In the following, 
we will label this setup as ATLAS-NF1.

\begin{figure}[ht]
  \centering
  \includegraphics[width=0.45\textwidth]{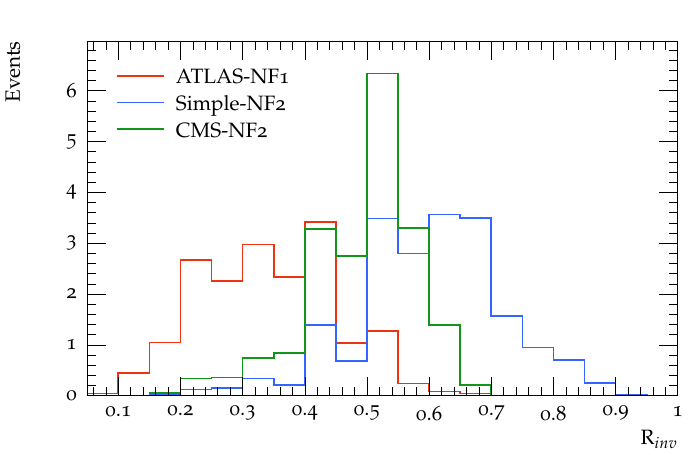}\qquad
  \includegraphics[width=0.48\textwidth]{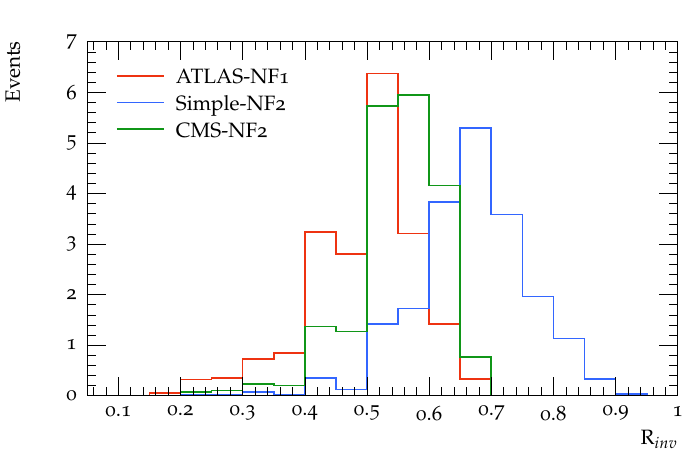}\\
  \includegraphics[width=0.45\textwidth]{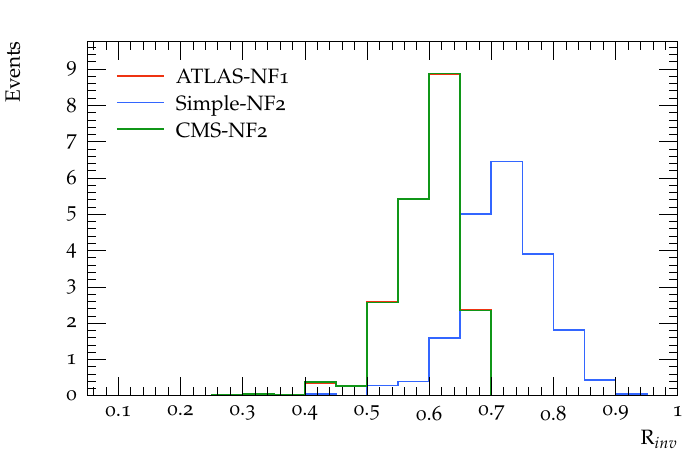}\qquad
  \includegraphics[width=0.45\textwidth]{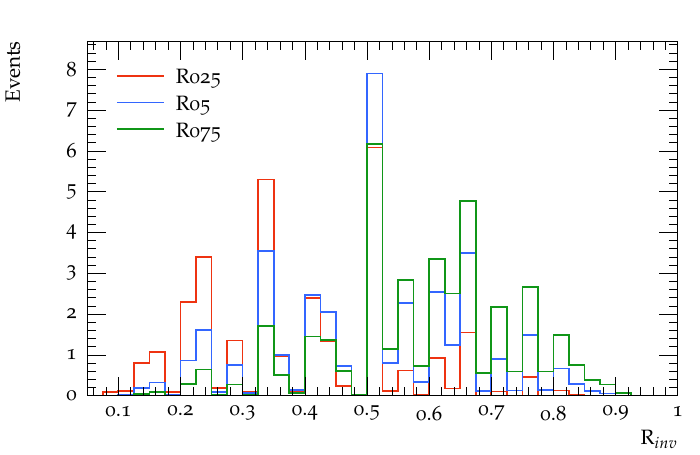}\\[1em]
  \caption{The output \Rinv\ calculated as the ratio of stable dark hadrons over 
  all dark hadrons shown for input \Rinv\ branching fraction of 0.25, 0.5 and 0.75 in the first three figures. ATLAS-NF1 indicates the $N_F=1$ set up in Ref.~\cite{ATLAS:2023swa}, CMS-NF2 indicates the $N_F=2$ set up in Ref.~\cite{CMS-EXO-19-020}, and the Simple-NF2 indicates the $N_F=2$ set up used here. The last figure shows what fraction of stable dark hadrons originate in the decay, for $N_F=2$.}
  \label{fig:rinvcomp}
\end{figure}

However, for larger values of $N_F$, if the additional off-diagonal stable dark hadrons are not subject to decay, the effective output \Rinv\, calculated as \RinvR\ is further increased. This non-conserving scenario, will be referred to as the Simple-NF2 setup, which is adopted in this study. Finally, the setup adapted in CMS search, will be referred to as CMS-NF2. In Fig.~\ref{fig:rinvcomp}, output \Rinv\ calculated by \RinvR\ approach is shown for different input \RinvBF\ values. It verifies the earlier assumption,
that the two definition indeed yield different results.
The figure also depicts the fraction of stable dark hadrons originating from decays, which remains independent of the input \RinvBF. This suggests that, in this case, a significant fraction of dark hadrons arise from the hadronisation process, implying that no definition of \Rinv\ can effectively probe a low effective value. This leaves open the question of probing the full range in an experimental search.

\begin{figure}[ht]
  \centering
  \includegraphics[width=0.45\textwidth]{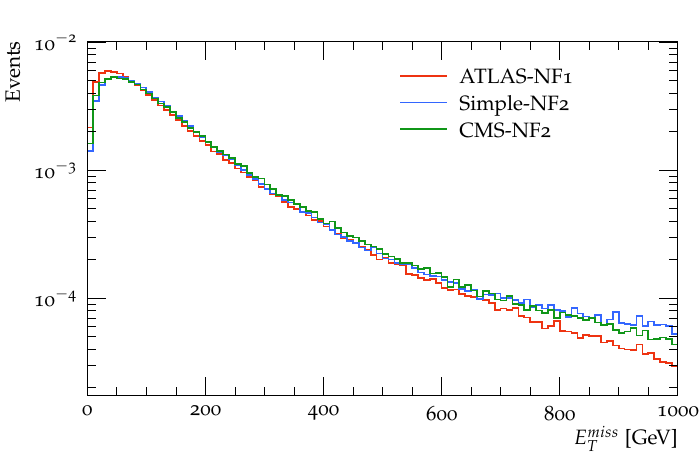}\qquad
  \includegraphics[width=0.48\textwidth]{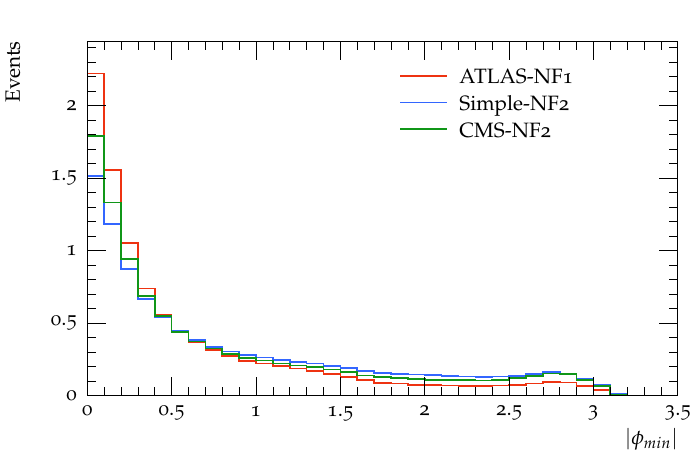}\\
  \includegraphics[width=0.45\textwidth]{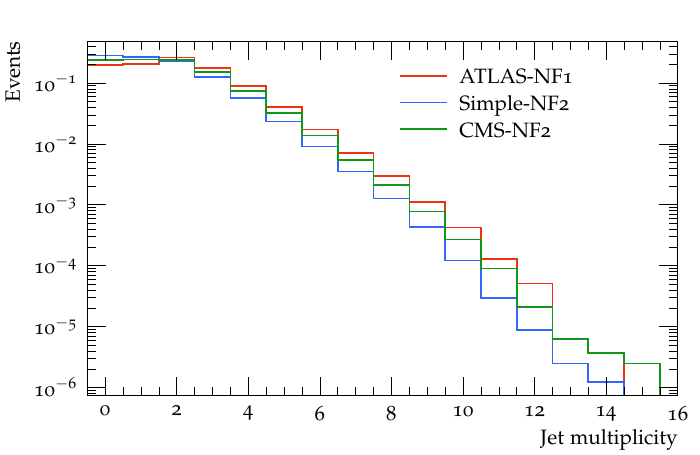}\qquad
  \includegraphics[width=0.45\textwidth]{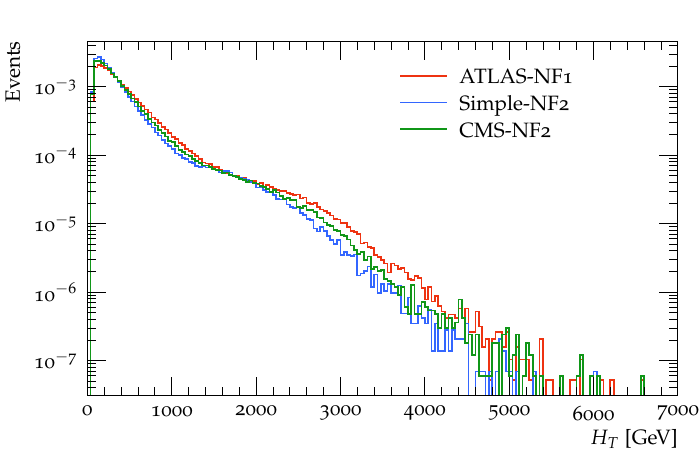}\\[1em]
  \caption{The area-normalised distributions of \met (top left), \minphi\ (top right), jet multiplicity (bottom left) and \HT (bottom right) for the three different approaches (corresponding to a mediator mass of 3000 GeV and \Rinv\ of 0.5) are shown. ATLAS-NF1 indicates
the $N_F=1$ set up in Ref.~\cite{ATLAS:2023swa}, CMS-NF2 indicates the $N_F=2$ set up in Ref.~\cite{CMS-EXO-19-020}, and the Simple-NF2 indicates the $N_F=2$ set up used here, without a complicated decay chain.}
  \label{fig:nfcomp}
\end{figure}

In Fig.~\ref{fig:nfcomp}, comparisons of the magnitude of missing transverse momentum (\met),
$\Delta \Phi$ of the jet closest to the direction of the missing transverse momentum (\minphi),
number of jets, and the scalar sum of \pT of the jets (\HT) are shown for these three different approaches for a fixed value of \Rinv\ of 0.5. While the generations with $N_F$ of two 
does produce more stable DM content, the kinematic trends are not very different. The other values of \Rinv\ show similar trends as well.

\end{document}